\documentclass[aps,prl,twocolumn,showpacs,superscriptaddress,floatfix]{revtex4-1}
\usepackage[utf8]{inputenc} 
\usepackage{amssymb,amsmath}
\usepackage{graphicx}
\usepackage{braket}
\usepackage{siunitx}
\usepackage{xcolor}
\usepackage{gensymb} 
\usepackage{bbold}
\usepackage{physics}

\def\lambdar{\lambda_{\rm R}}                           
\def\kr{k_{\rm R}}                            			
\def\Er{E_{\rm R}}                            			
\def\Rb87{^{87}\mathrm{Rb}}                             
\def\K40{^{40}\mathrm{K}}                    		    

\def\ex{\mathbf{e}_x}  
  
\def\ey{\mathbf{e}_y}  
\def\ez{\mathbf{e}_z}

\newcommand{\Fx}{\langle \hat{F}_x \rangle}

\newcommand{\rfph}{\phi_{\text{rf}}}
\newcommand{\Omrf}{\Omega_{\text{rf}}}

\definecolor{green}{rgb}{0.0,0.5,0.0}
\usepackage[normalem]{ulem}

\begin{document}

\title{Floquet engineering topological Dirac bands}
\author{Mingwu Lu}
\author{G.~H.~Reid}
\author{A.~R.~Fritsch}
\author{A.~M.~Pi\~{n}eiro}
\author{I.~B.~Spielman}
\affiliation{Joint Quantum Institute, National Institute of Standards and Technology, and University of Maryland, Gaithersburg, Maryland, 20899, USA}
\email{spielman@nist.gov}
\date{\today}

\begin{abstract}
We experimentally realized a time-periodically modulated 1D lattice for ultracold atoms featuring a pair of linear bands, each associated with a Floquet winding number: a topological invariant.
These bands are spin-momentum locked and almost perfectly linear everywhere in the Brillouin zone (BZ), making this system a near-ideal realization of the 1D Dirac Hamiltonian.
We characterized the Floquet winding number using a form of quantum state tomography, covering the BZ and following the micromotion through one Floquet period.
Lastly, we altered the modulation timing to lift the topological protection, opening a gap at the Dirac point that grew in proportion to the deviation from the topological configuration.
\end{abstract}

\maketitle

Topologically protected edge modes are present in far-ranging systems from 2D and 4D quantum Hall systems~\cite{Klitzing1980, Zhang2001}, $Z_2$ topological insulators~\cite{Hasan2010}, to atmospheric waves~\cite{Delplace2017}.
Systems with time-periodic driving, described by Floquet theory, allow for new topological invariants~\cite{Kitagawa2010,lindner2011floquet,Rudner2013a} including the Floquet winding number, leading to new protected quantities.
Here we study atomic Bose-Einstein condensates (BECs) in the lowest two bands of a periodically driven 1D optical lattice and observe a pair of protected chiral bands that are a near-ideal realization of the 1D Dirac Hamiltonian.
We directly extract the topological winding number from the time-resolved micromotion and find that altering the modulation timing opens a gap at the Dirac point.

The conventional bulk-edge correspondence yields protected edge bands that reside on the system's surface and therefore have lower dimension than the bulk.
For example in 2D, $Z_2$ topological insulators have a pair of counter-propagating spin-momentum locked 1D edge modes.
By contrast we observe 1D topologically protected bands derived from a periodically driven 1D system, where the topological protection results from a non-zero Floquet winding number~\cite{Kitagawa2010} defined in terms of the 1+1D space defined by crystal momentum $q$ and time $t$~\cite{budich2017helical}.
These bands are spin-momentum locked, intersect at $q=0$ and have the remarkable property of being linear everywhere in the Brillouin zone (BZ), which is inconsistent with the usual requirement that bands be continuous (and differentiable) as they cross the edge of the BZ.
The periodic quasienergy structure of Floquet systems allows these bands to smoothly cross the edge of the BZ by entering the next quasienergy zone.

\begin{figure}[t]
\includegraphics{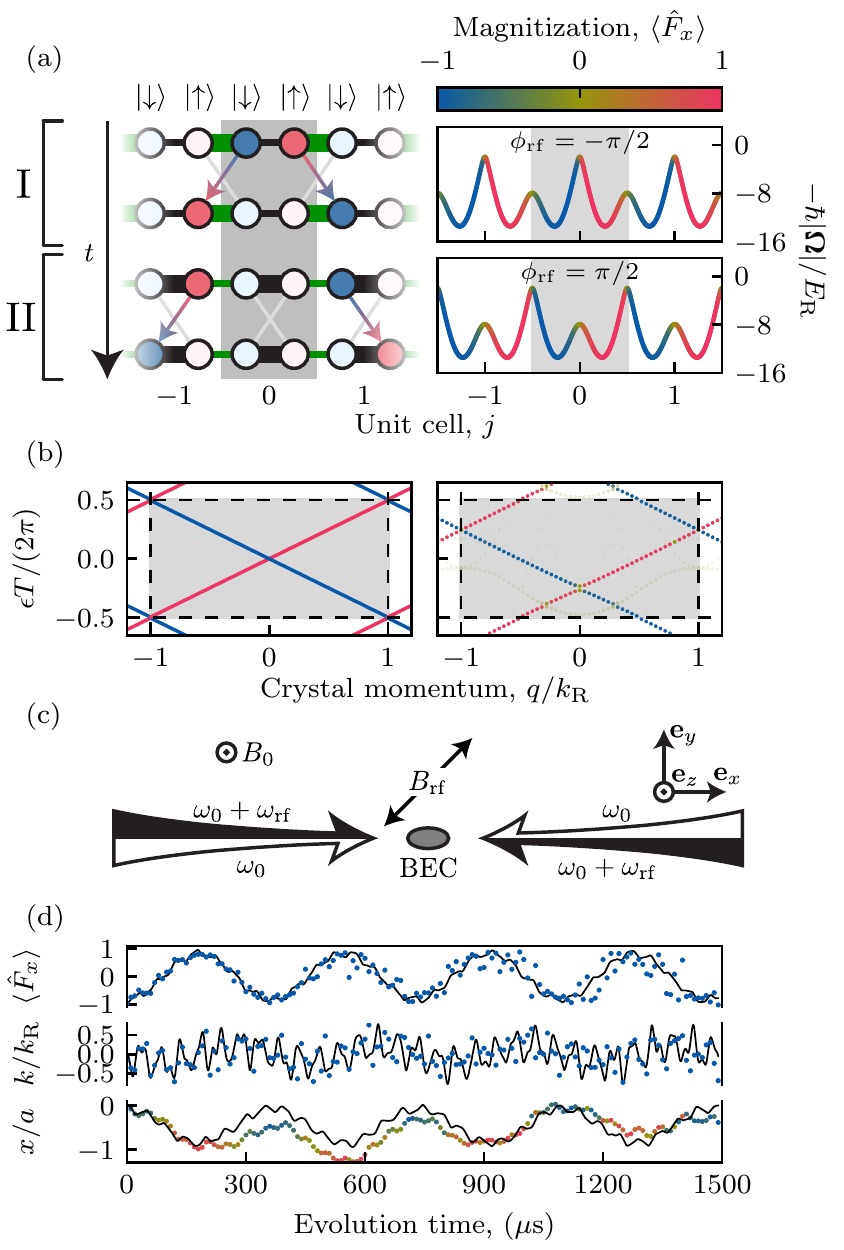}
\caption{Concept.
(a) Left: Switching protocol, with the $j=0$ unit cell marked in grey.
In configuration I atoms tunnel between neighboring unit cells (bold green links); in configuration II, they tunnel within the same unit cell (bold black links).
Right: spin-dependent lattice potential implemented for each configuration, colored by $\Fx$.
(b) Floquet band structure colored according to $\Fx$ sampled stroboscopically.
The lowest BZ and quasienergy zone is marked in grey.
Left: idealized SSH model; right: numerical lattice model [see Ref.~\cite{SeeSM} the supplementary material (SM)] computed for $\hbar[\Omrf, \Omega_+, \Omega_-] = [4.48, 12.5, 5.69]E_{\rm R}$.
(c) Schematic.
The BEC was illuminated by counter-propagating Raman lasers and an rf magnetic field. 
(d) Static lattice tunneling with data (markers) and simulations (black curves).
Upper: magnetization $\Fx$.
Middle: group velocity.
Bottom: displacement computed from integrated group velocity, and colored according to $\langle\hat F_x(t)\rangle$ using the color scale in (a). 
\label{Fig_Setup}}
\end{figure}

Remarkably all of these features are present in a periodically modulated Su-Schrieffer-Heeger (SSH) model~\cite{su1979solitons}
\begin{align*}
    \hat{H} &= -\sum_j \left[J\dyad{j+1, \downarrow}{j, \uparrow} + J^{\prime} \dyad{j, \downarrow}{j, \uparrow}+ \text{h.c.}
    \right]
\end{align*}
that approximates our 1D bipartite lattice~\cite{lu2016geometrical}.
Figure~\ref{Fig_Setup}(a) shows that each unit cell (labeled by integer $j$) consists sites that we denote by $\ket{\uparrow}$ and $ \ket{\downarrow}$ to emphasize their role as a pseudospin degree of freedom.
$J'$ and $J$ are the tunneling strengths within a unit cell and between adjacent unit cells, respectively.


Following Ref.~\onlinecite{budich2017helical}, we implemented a Floquet ``switching'' protocol where the lattice periodically alternates between a ``configuration I'' with $J^\prime \approx 0$ and $J = J_0$ and a reversed ``configuration II'' with period $T$.
This allows intercell tunneling $\ket{j+1, \downarrow } \leftrightarrow \ket{j, \uparrow}$ during the first half period and intracell tunneling $\ket{j, \downarrow} \leftrightarrow \ket{j, \uparrow}$ during the second half period.
When $J_0 T = \pi$ each half period implements a $\pi$-pulse, exchanging the amplitude between sites.
Figure~\ref{Fig_Setup}(a) conceptually illustrates how this exact timing leads to a displacement of one unit cell per Floquet period, with $\ket{j,\uparrow}\rightarrow\ket{j+1,\uparrow}$ and  $\ket{j,\downarrow}\rightarrow\ket{j-1,\downarrow}$.
This gives constant velocity $v = \pm a/T$, (pseudo)spin-momentum locked motion under stroboscopic observation, where $a$ is the lattice constant.
Together these features are captured by a 1D Floquet Dirac Hamiltonian
\begin{align}
    \hat{H}^{\text{F}}(q) &= q v \hat{\sigma}_z\label{eq:dirac}
\end{align}
describing massless (i.e. gapless) relativistic particles. 
Any deviation from exact $\pi$-pulses opens gaps in the quasienergy spectrum $\hbar\epsilon_\alpha(q)$, where $\alpha$ labels the quasienergy band.
For each initial pseudospin, different crystal momentum states start and end each driving period at the same point on the Bloch sphere, but follow different trajectories within the driving period.
We show that, taken over the whole BZ, these trajectories cover the Bloch sphere, giving winding numbers of $\pm1$ for initial pseudospins $\ket{\uparrow\downarrow}$.
A related experiment in a small synthetic dimension chain observed the drift of initially localized states~\cite{xiao2020periodic} but not the linear drift of crystal momentum eigenstates nor the band topology.

We observed these properties using a $\Rb87$ BEC in a 1D bipartite optical lattice, resulting from an effective magnetic field ${\boldsymbol \Omega}(\hat{x})$ due to a combination of ``vector'' light shifts and an rf magnetic field~\cite{lu2016geometrical}.
Figure ~\ref{Fig_Setup}(a, right) shows the lowest energy adiabatic potential of our lattice in its I and II configurations.


{\it Experiments} Our experiments began with small $N\approx 10^4$ atom~\footnote{These small numbers decreased the atomic density and limited unwanted scattering processes~\cite{Campbell2006}} $^{87}$Rb BECs in a crossed optical dipole trap (ODT) 
in the $\ket{f=1,m_F=-1}$ hyperfine ground state.
The ODT, formed by two intersecting $1064\ {\rm nm}$ laser beams traveling along $\ex$ and $\ey$, had trap frequencies $(\omega_x, \omega_y, \omega_z)/2\pi \approx (15, 150, 100)\ {\rm Hz}$.
A bias magnetic field $B_0\approx 0.1\ {\rm mT}$ Zeeman-split the three $m_F$ states by $\omega_{\rm Z} / 2\pi \approx 1\ {\rm MHz}$.
These states were dressed by a radiofrequency (rf) magnetic field with frequency $\omega_{\rm rf}$ and two laser beams counterpropagating along $\ex$ driving Raman transitions.
As shown in Fig.~\ref{Fig_Setup}(c), each Raman beam had frequency components $\omega_0$ and $\omega_0 + \omega_{\rm rf}$; $\rfph$ denotes the relative phase between the rf field and the Raman beat tone.
The wavelength $\lambdar=2\pi c/\omega_0 = 790.03(2)\ \text{nm}$ of the Raman lasers~\footnote{This wavelength serves to nearly minimize the spontaneous emission rate at fixed  Raman coupling strength as well as cancel the state-independent scalar light shift.  
All uncertainties herein reflect the uncorrelated combination of single-sigma statistical and systematic uncertainties.} defines the single-photon recoil wave-vector $\kr= 2 \pi /\lambdar$ and energy $E_{\rm R} = \hbar^2 \kr^2 / 2 m$, with speed of light $c$ and reduced Planck constant $\hbar$.
The atoms interact with these fields via a Zeeman like Hamiltonian~\cite{juzeliunas2012flux} $\hat H_{\rm int}\!=\!{\boldsymbol \Omega}({\hat x})\!\cdot\! \hat {\bf F}$, with total atomic angular momentum operator $\hat {\bf F}$.
The effective magnetic field ${\boldsymbol \Omega}({\hat x}) = [\Omrf \cos(\rfph) + \bar{\Omega} \cos(2 \kr \hat{x}), -\Omrf \sin(\rfph) - \delta \Omega \sin(2 \kr \hat{x}) , \sqrt{2} \delta]/\sqrt{2}$ is defined in terms of the detuning $\delta = \omega_{\rm Z} - \omega_{\rm rf}$; the  rf coupling strength $\Omrf$; and $\bar{\Omega} = \Omega_+ + \Omega_-$ and $\delta \Omega = \Omega_+ - \Omega_-$, derived from the two Raman coupling strengths $\Omega_{\pm}$.
The lowest energy adiabatic potential formed a spin-dependent bipartite lattice~\cite{lu2016geometrical}, shown for two choices of $\rfph$ in Fig.~\ref{Fig_Setup}(a).
As indicated by the magnetization of the adiabatic potentials, the $\ket{\uparrow,\downarrow}$ sites are highly spin polarized, corresponding to atomic states $\ket{m_x = \pm1}$.
The potential minima are degenerate for $\rfph = \mp \pi /2$, where $\mp$ selects between configurations I and II.
All other values of $\rfph$ introduce an energy difference $\Delta$ between $\ket{\uparrow,\downarrow}$ that, while absent in the SSH model, is useful for state preparation~\cite{SeeSM} and readout.

Following all our experiments, we measured the spin-resolved momentum distribution by first removing the the Raman lasers and the rf field.
An rf pulse induced a $\pi/2$ rotation around $\hat F_y$, transforming eigenstates of $\hat F_x$ to our $\hat F_z$ measurement basis; we then initiated time of flight (ToF) by extinguishing the trapping lasers.
The atoms then evolved for $12\ \text{ms}$ during which time a magnetic field gradient along $\ey$ Stern-Gerlach separated the three $m_F$ states; after this ToF, the density distribution was detected by resonant absorption imaging.
This allowed us to separately infer the overall populations in the $\ket{\uparrow}, \ket{\downarrow}$ sites.


\begin{figure}
\includegraphics{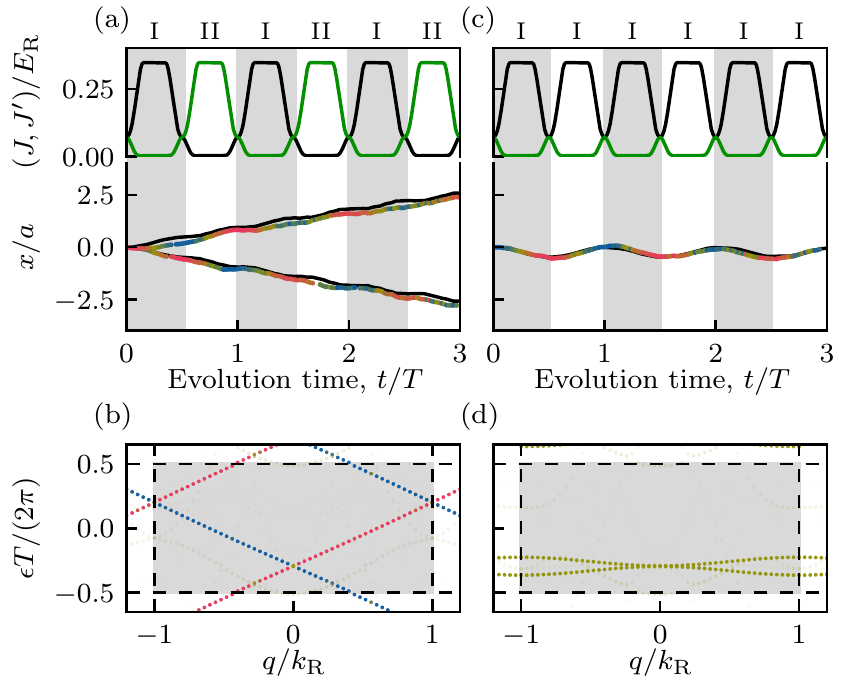}
\caption{
Floquet protocols. 
(a,b) switching protocol and (c,d) the single-configuration protocol.
(a,c) Computed intercell (black) and intracell (green) tunneling strengths and displacement (colored according $\Fx$).
Grey and white bands indicate the different configurations.
(b,d) Floquet quasienergies [using the same color scale as in Fig.~\ref{Fig_Setup}(a)].
}
\label{Fig_Linear}
\end{figure}


Our procedure for loading BECs into the bipartite lattice with occupation on sites $\ket{\uparrow}$ or $\ket{\downarrow}$ began with $\rfph = 0$ or $\pi$ to select which state is loaded, at which time the coupling fields and detuning were adiabatically ramped to their final values in $2.5\ \text{ms}$.
Lastly, we selected between configuration I and II by abruptly changing $\rfph$ to $\mp \pi /2$.
The resulting $q=0$ pseudospin polarized state was an equal superposition of our lattice's lowest two bands; following loading, atoms resonantly tunneled between the strongly coupled neighboring pseudospins~\cite{Trotzky2008,Brown2015}.


{\it Dispersion } Figure~\ref{Fig_Setup}(d) plots this tunneling in configuration I for atoms prepared in $\ket{\downarrow}$ where data is plotted by markers and the solid curves are the results of our numerical model.
The top panel shows the measured magnetization $\Fx$ coherently oscillating with $366(3)\ \mu {\rm s}$ period, resulting from motion between neighboring sites.
We separately observe near-zero population in $\ket{m_x=0}$ during this evolution, enabling the mapping $\ket{\uparrow,\downarrow}\rightarrow\ket{m_x=\pm 1}$.
The scatter increases at long times, indicating the onset of dephasing, likely from a combination of optical path changes from acoustic vibrations, laser intensity noise, and magnetic field instabilities.

Figure~\ref{Fig_Setup}(d, middle) plots the instantaneous group velocity obtained from the momentum distribution measured in ToF~\cite{Lin2011a}.
The high frequency oscillations are repeatable and have amplitude consistent with the $\approx 7\ \%$ occupation of higher bands anticipated by our numerical modeling (black).
The bottom panel integrates the group velocity~\cite{stuhl2015visualizing}, giving the BEC's displacement as it tunnel-oscillates between adjacent lattice sites separated by nearly $1/2$ of a unit cell, $\approx 200\ {\rm nm}$.
While the higher frequency components are conspicuous in group velocity, they play little role in atomic displacement at the tunneling timescale, since integration acts as a low-pass filter that suppresses these components.

Having demonstrated the behavior of the static lattice, Fig.~\ref{Fig_Linear}(a) depicts the configuration switching protocol with near optimal timing.
This was achieved by suddenly changing the phase $\rfph$, ideally every half tunneling period as evoked in Fig.~\ref{Fig_Setup}(a).
To avoid exciting higher bands with these abrupt switches, we smoothly ramped $\Omrf$ to zero, changed $\rfph$, and reversed the ramp, smoothly changing $J$ and $J^\prime$ as in the Fig.~\ref{Fig_Linear}(a, top).
The drive period $T= 448\ \mu{\rm s}$ increased from the $\approx 366\ \mu{\rm s}$ bare tunneling period [Fig.~\ref{Fig_Setup}(d)], a slow-down resulting from the time spent ramping $\Omrf$ to and from zero, during which time tunneling was suppressed.
We empirically found the rf phases to achieve configurations I and II differed by $\rfph^{({\rm II})} - \rfph^{({\rm I})} \approx 1.03\pi$ rather than $\pi$ as predicted by our model.
In addition, we observed a $6(2)\ \%$ difference in their tunneling periods~\footnote{These observations cannot be explained by our model, indicating the presence of distortions of the lattice potentials; we find that a weak optical lattice, for example from small retro-reflections of our Raman lasers, can explain both of these observations.}.
We compensated for this in our modulation scheme by reducing the time spent in configuration II proportionally.

Fig.~\ref{Fig_Linear}(a, bottom)\ shows results for atoms prepared in $\ket{\uparrow}$ (positive slope) and $\ket{\downarrow}$ (negative slope).
Following each Floquet period, the magnetization of both trajectories (indicated by the color of the markers), returned to their initial values, demonstrating spin-momentum locking of Floquet eigenstates.
These data show a near-linear increase of displacement sustained over many Floquet periods consistent with our numerically modeled time evolution (black), yielding drift velocities $\pm 0.89(4) a/T$ and $\pm 0.86(2) a/T$, respectively.
These differ from the ideal drift velocity $a/T$, i.e., one unit cell per cycle that seems apparent in the band structure [Fig.~\ref{Fig_Linear}(b)].
Our numerics indicate this results from the nonzero value of both $J$ and $J^\prime$ during our rf-switching stage [black and green curves in Fig.~\ref{Fig_Linear}(a, top)], allowing unwanted tunneling; and the departure of our physical system from the tight-binding SSH model.

To confirm the importance of the configuration-switching protocol, we introduced a single-configuration protocol with the same $\Omrf$ ramps but with constant $\rfph$ [Fig.~\ref{Fig_Linear}(c)]. 
The displacement and magnetization measured following this protocol are oscillatory and correspond to tunneling confined within a single double well.
Figure~\ref{Fig_Linear}(d) shows the associated Floquet band structure with a quadratic touching point, reminiscent to those in bilayer graphene~\cite{Zhang2009a}.
The curvature of these bands results from the same deviations from the idealized switching protocol that lead to differences from the expected drift velocity in the configuration-switching protocol. 


{\it Winding number} Similar to adiabatic charge pumps~\cite{thouless1983quantization}, the topology of 1D Floquet bands is characterized by an integer valued winding number
\begin{align}
\nu &= \frac{1}{2\pi} \int_{\rm BZ} \int_0^T dq dt F(q, t),\label{eq:winding}
\end{align}
defined in terms of the Berry curvature $F(q,t) = (\braket{\partial_q \psi(q,t)}{\partial_t \psi(q,t)}-{\rm c.c.})/i$.
We reconstruct the two component (pseudo-)spinor $\ket{\psi(q,t)}$ for all crystal momentum states over one period of modulation using quantum state tomography~\footnote{Quantum state tomography recovers the density operator, and we take its principal eigenvector as $\ket{\psi(q,t)}$.
This leaves the overall wavefunction phase undetermined. Due to the gauge invariance of $F(q,t)$ this phase does not affect the value of $\nu$.} and directly compute $\nu$~\cite{Alba2011} using Ref.~\onlinecite{fukui2005chern} for discretely sampled data.

\begin{figure}
\includegraphics{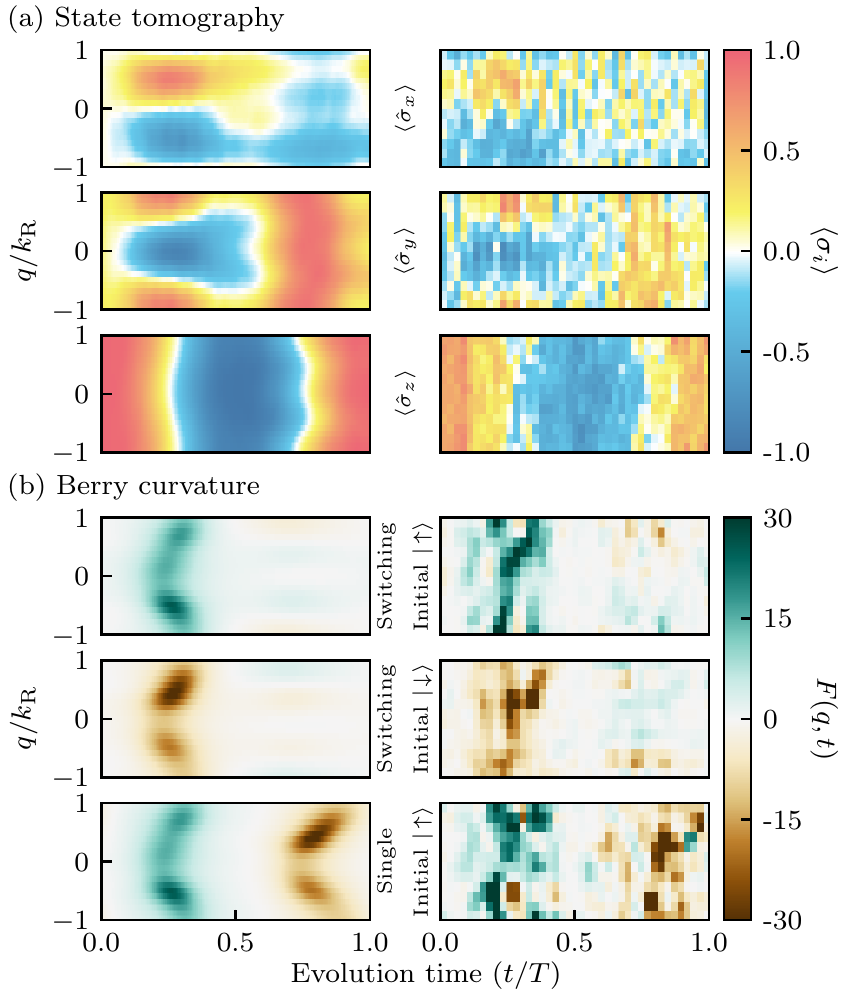}
\caption{Crystal momentum resolved pseudospin micromotion and corresponding Berry curvature. Upper three panels: Numerical model (left) and unfiltered experimental data (right) for pseudospin components for configuration switching protocol (initial states $\ket{\uparrow}$). Lower three panels: Berry curvature based on filtered experimental data (right) and numerical simulation (left) for configuration switching protocol with initial states $\ket{\uparrow}$ and $\ket{\downarrow}$ and single-configuration protocol (initial state $\ket{\uparrow}$).
} 
\label{Fig_Pseudospin}
\end{figure}

Our standard measurement gives the population in the $\{ \ket{\uparrow}, \ket{\downarrow} \}$ states from which we obtain $\expval{\hat\sigma_z}$.
To measure $\expval{\hat\sigma_x}$ and $\expval{\hat\sigma_y}$, we designed lattice configurations for which unitary evolution implemented pseudospin rotations generated by $(\hat \sigma_x + \hat\sigma_z)/\sqrt{2}$ and $\hat \sigma_x$, respectively~\cite{Reid2021}.
We applied these operations after the system evolved for a Floquet time $t$ and parallelized the measurement by filling the ground band of our initial lattice~\cite{pineiro2019sauter} to measure all $q$ states simultaneously~\cite{Valdes-Curiel2021}~\footnote{To aid in filling the band, we increased the longitudinal trap frequency to $\omega_x/ 2 \pi \approx 25\ {\rm Hz}$ for experiments discussed in this section.}.
These data yielded the crystal momentum resolved pseudospin magnetization ${\bf m}(q,t) = (\expval{\hat\sigma_x(q,t)},\expval{\hat\sigma_y(q,t)},\expval{\hat\sigma_z(q,t)})$ from the measured populations following each rotation.
Figure~\ref{Fig_Pseudospin}(a) plots all three components of ${\bf m}(q,t)$ for a single Floquet cycle of our configuration-switching protocol, starting in state $\ket{\uparrow}$.
The left panels show the result of our numerical simulation; the experimental data on the right is consistent with the simulations.

Our measurement of ${\bf m}(q,t)$ suffices to obtain the associated Floquet winding number~\cite{Kitagawa2010} using Eq.~\eqref{eq:winding}.
Evaluating the Berry curvature requires differentiation of noisy data, so we applied a low-pass Gaussian filter (with root mean squared widths $\Delta t = 10\ \mu{\rm s}$ and $\Delta q = \kr/6$) prior to computing $F(q,t)$.
Panel (b) plots the resulting Berry curvatures $F(q,t)$ for our configuration switching protocol with initial states $\ket{\uparrow}$ and $\ket{\downarrow}$, as well as our single-configuration protocol (top, middle and bottom respectively).
For $\ket{\uparrow}$, $F(q,t)$ has a net positive contribution for $t < T/2$; while for $t > T/2$ both positive and negative structures are present; these cancel upon integration.
All together we find $\nu_{\uparrow,\downarrow} = \left\{0.991(5), -0.998(4)\right\}$ for systems initialized in $\ket{\uparrow}$ or $\ket{\downarrow}$; this is in very good agreement with $\left\{0.9994, -0.9995\right\}$ obtained by performing the same analysis on numerically simulated data.
Uncertainties in our lattice parameters (leading to deviations from optimal timing) and imperfect state preparation can cause the time-evolution to be not perfectly $T$-periodic, yielding non-integer $\nu$ even without the technical noise present in experiment.
For comparison, panel (b) bottom shows $F(q,t)$ for our single-configuration protocol, for which we obtain $\nu=0.01(2)$, compared to $\nu=0.0019$ from simulation. 
Here our initial state was fully magnetized, an eigenstate of the ideal switching protocol, but a coherent superposition of the two bands shown in Fig.~\ref{Fig_Linear}(d).

Unlike topological invariants in static systems, the Floquet winding number is directly linked to $\epsilon_\alpha(q)$ via~\cite{Kitagawa2010}
\begin{align}
\nu &=  \sum_\alpha \left[\frac{T}{2\pi}\int_{\rm BZ} dq \partial_q \epsilon_\alpha(q)\right]. \label{eq:winding_energy}
\end{align}
Each term of the sum measures the difference in quasienergy at the $\pm$ edges of the BZ for the $\alpha$th band; the integral is zero for bands that link at the edge of the BZ (such as our single-configuration protocol) since $\epsilon_\alpha(-\kr)=\epsilon_\alpha(\kr)$.
By contrast our Dirac-like bands change in quasienergy by $\pm 2\pi/T$, each contributing $\pm 1$ to the sum suggesting $\nu=0$.
Our configuration switching protocol obeys a chiral symmetry~\cite{budich2017helical}, for example the symmetry operation $\hat \Sigma = \hat\sigma_x$ takes $\hat \Sigma^\dagger \hat H^{\rm F} \hat \Sigma = -\hat H^{\rm F}$, thereby separating state-space into decoupled $\uparrow$ and $\downarrow$ subspaces.
Individually these have $\nu_{\uparrow\downarrow} = \pm 1$.


\begin{figure}[b]
\includegraphics{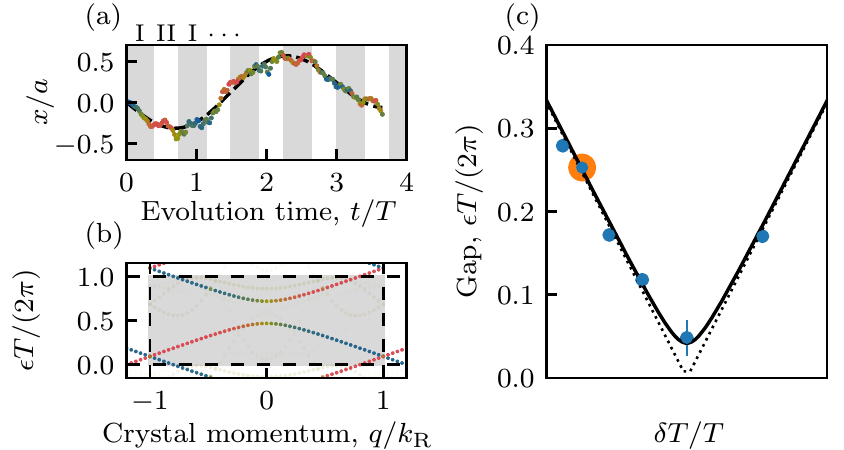}
\caption{
Breaking of chiral symmetry.
(a) Time evolution with Floquet period $T = 330\ \mu{\rm s}$, away from the optimal point $T_0 = 438\ \mu{\rm s}$, colored according to the instantaneous magnetization using the color scale in Fig~\ref{Fig_Setup}.
Configurations (grey rectangles) are plotted along with the data.
(b) Computed spectrum for data in (a), and circled in orange in (c).
(c) Zitterbewegung frequency as a function of $T$ showing the gap closing at the symmetry point.
}
\label{fig4}
\end{figure}

{\it Fine-tuning} The chiral symmetry is present only for a fine-tuned switching protocol, i.e., tunneling $\pi$ pulses as discussed in the context of Eq.~\eqref{eq:dirac}; for example, changing the tunneling period to $T+\delta T$ open a gap $\approx 2 J_0 \abs{\delta T} / T$ in the Floquet spectra at the center of the BZ~\footnote{$H_2$ in Equation~(2) of Ref.~\onlinecite{budich2017helical} differs by a factor of $-1$ from the SSH model; this in effect swaps the edge and center of the BZ.} leading to non-topological bands with massive Dirac dispersion.

Figure~\ref{fig4}(a) plots the time evolving position when $J_0 T < \pi$; the data is colored according to its instantaneous magnetization and the gray boxes mark the configurations.
This shows the first switch occurring before the magnetization inverts, and at longer timescales the position undergoes periodic oscillations---zitterbewegung~\cite{leblanc2013zitter}---arising from the quantum interference~\footnote{This interference arises because our initial state (aligned along the pseudo-spin $\ez$ direction) is an equal superposition of the gapped eigenstates (aligned in the pseudo-spin $\ex$-$\ey$ plane).} of the two gapped bands at $q=0$, shown in Fig.~\ref{fig4}(b).  

The dependence of the gap on $\delta T$ in Fig.~\ref{fig4}(c) is in near perfect agreement with the simple model (dashed lines), and fitting to a hyperbola provides an estimate $0.05(1)\times(2\pi/T)$ of the gap in our fine-tuned configuration, indicating that our experiment was very close to the optimal configuration.



{\it Discussion and outlook} Topological systems can be organized by their symmetries~\cite{Kitaev2009}, and the breaking of the chiral symmetry of our system is similar to $Z_2$ topological insulators where any small magnetic field breaks time reversal symmetry and opens a gap where the edge bands cross.

Our protocol realizes a diabatic quantized charge pump, complementing topological and geometrical Thouless pumps~\cite{thouless1983quantization} realized with ultracold atoms~\cite{lohse2016thouless,nakajima2016topological,lu2016geometrical}.
Adiabatic Thouless pumps are also characterized by the Floquet topological index in Eqs.~\eqref{eq:winding} and \eqref{eq:winding_energy}.
Similarly the (nearly) adiabatic Floquet time evolution operator factorizes into decoupled subspaces (not labeled by $\ket{\uparrow\downarrow}$). 
At any finite drive frequency the evolution operator mixes these subspaces resulting in topologically trivial bands.
As a result, Thouless pumps do not continuously connect to the diabatic case discussed here; in addition, our control trajectory directly traverses the gap-closing point in the SSH model (when $J=J^\prime$) and thus could not operate as an adiabatic pump.

Analogous switching schemes can create topological edge~\cite{Kitagawa2010} and surface states~\cite{Huang2021a} in 2D and 3D, which unlike our 1D system, are similar to conventional topological systems with an insulating bulk and dispersing edge modes and are related to a recently observed anomalous 2D Floquet system~\cite{Wintersperger2020}.

\begin{acknowledgments}
The authors thank W.~D.~Phillips for productive discussions, and C.~W.~Clark and C.~A.~Bracamontes for carefully reading the manuscript.
This work was partially supported by the National Institute of Standards and Technology, and the National Science Foundation through the Physics Frontier Center at the Joint Quantum Institute.
\end{acknowledgments} 

\bibliography{main}

\newpage

\clearpage
%
\widetext
\begin{center}
\textbf{\large Supplemental Materials for ``Floquet engineering topological Dirac bands''}
\end{center}

\section{Interactions}

We estimate the strength of interactions by modeling the Wannier orbitals of our deep lattice as Gaussian, resulting from a harmonic expansion of the minima of the adiabatic potential.
The local oscillator frequency is largest for $\Omrf=0$, so to bound the interaction strength we study this configuration.
Since $\bar\Omega > \delta\Omega$ the potential minima are located at $x = j \lambda / 2$ for integer $j$.  

We start with harmonic expansion of a simple cosinusoidal optical lattice of depth $s$
\begin{align*}
V(x) &\approx \frac{1}{2} \left(2 s \Er \kr ^2\right) x^2
\end{align*}
that gives an oscillator frequency $\omega_{\rm ho} = 2 \Er \sqrt{s}/\hbar$ and oscillator length $\ell_{\rm ho} = (\kr s^{1/4})^{-1}$.
An analogous expansion for our adiabatic potential gives an effective depth
\begin{align*}
s_{\rm eff} &= \sqrt{2} \frac{(\bar\Omega/\Er + \delta\Omega/\Er)(\bar\Omega/\Er - \delta\Omega/\Er)}{\bar\Omega/\Er},
\end{align*}
which for our experimental parameters is $s_{\rm eff}\approx 22$

We find the fractional change in the total energy when initially uniformly distributed atoms in every $\lambda/2$ range are compressed into a Gaussian wavepackets to be
\begin{align}
\frac{\mu}{\mu_0} &= \sqrt{\frac{\pi}{2}} s^{1/4}
\end{align}
which for our parameters is about $2.7$ .
This change in chemical potential can be attributed to an effective interaction strength changed by the ratio $\mu/\mu_0$.

For our $\approx 10^4$ atom $\Rb87$ BECs in their $(\omega_x, \omega_y, \omega_z)/2\pi \approx (15, 150, 100)\ {\rm Hz}$ harmonic trap, we obtain a chemical potential $\mu=h\times 390\ {\rm Hz}$, and incorporating the effective interaction strength leads to an effective chemical potential $\mu_{\rm eff} = h\times 570\ {\rm Hz}$.
This energy provides a rough estimate of a several milliseconds as the timescale for the onset of interaction induced dephasing effects, which is comparable to the dephasing times observed in experiment.  

\section{Numerical model}

We here introduce our numerical model describing non-interacting atoms subject to the combined Raman and rf coupling. 
The large trapping period of $60\ {\rm ms}$ along the lattice direction (which would be further increased by effective mass contributions) greatly exceeds the typical milliseconds time of interest for the experiments.
As a result we focus on a momentum-space description that naturally describes the light-matter interactions.

The coupling terms of the Hamiltonian have two contributions.
The rf field couples the internal states of the ground hyperfine manifold of $^{87}$Rb:  $\ket{f=1,m_F=-1}, \ket{f=1,m_F=0}, \ket{f=1,m_F=1}$ (abbreviated as $\ket{-1}, \ket{0}, \ket{1}$); the Raman interaction couples different momentum states while changing the atomic spin: $\ket{q, m_F} \leftrightarrow \ket{q \pm2\kr, m_F \pm1}$ for the two pairs of counter-propagating Raman beams with strengths $\Omega_+$ and $\Omega_-$, respectively.
These two couplings oscillate at $f=1\ \text{MHz}$, resonant with the linear Zeeman energy splitting and have a well defined relative phase of $\rfph = \varphi_{\text{rf}} - \varphi_{\text{ra}}$.

In the rotating frame and under the rotating wave approximation, the full Hamiltonian takes a block diagonal form
\begin{equation}
\mathbf{H} =
    \begin{pmatrix}
    \ddots & \mathbf{C} & &\\
     \mathbf{C}^\dagger & (\mathbf{A}_{N-1} + \mathbf{B}) & \mathbf{C} &  \\
     & \mathbf{C}^\dagger & (\mathbf{A}_N+\mathbf{B}) & \mathbf{C} \\
     & & \mathbf{C}^\dagger & (\mathbf{A}_{N+1}+\textbf{B}) & \mathbf{C} \\
     & & & \mathbf{C}^\dagger & \ddots
    \end{pmatrix}
\end{equation}
in the basis $\{\ket{q + 2N\kr, 1},\ket{q + 2N\kr, 0}, \ket{q + 2N\kr, -1} \}_{ N \in \mathcal{Z}}$.
The matrix
\begin{equation}
\mathbf{A}_N =\hbar^2(q-2N\kr)^2/2m \times \mathbf{1}
\end{equation}
describes the kinetic energy,
\begin{equation}
\mathbf{B} =
    \begin{pmatrix}
     \delta & \Omrf \text{exp}(-i\rfph)/2 & 0 \\
    \Omrf \text{exp}(i\rfph)/2  &  -\varepsilon & \Omrf \text{exp}(-i\rfph)/2 \\
    0 & \Omrf \text{exp}(i\rfph)/2 & -\delta
    \end{pmatrix}
\end{equation}
describes the rf coupling, detuning and quadratic Zeeman shift (with strengths $\Omrf$, $\delta$, and $\varepsilon$) and
\begin{equation}
\mathbf{C} = \frac{1}{2}
    \begin{pmatrix}
    0 & \Omega_+ & 0 \\
    \Omega_- & 0 & \Omega_+ \\
    0 & \Omega_- & 0\\
    \end{pmatrix}.
\end{equation}
describes the Raman coupling. 

In the experiment the detuning was stabilized near zero, with $|\delta|< 0.2\ \Er$ by monitoring the resonance condition via microwave sampling technique~\cite{lu2016geometrical} and tuning the bias field accordingly.
The quadratic Zeeman shift $\epsilon = 0.04\ \Er$ was small, but not ignorable.

In a periodically driven system, the Floquet Hamiltonian $\hat{H}_F$ can be defined through the relation $\hat{U}(t + T,t) = \text{exp}(-i T \hat H^{\rm F}_t / \hbar)$. In experiment, we always choose the initial time $t= 0$ so the Floquet Hamiltonian is uniquely defined, so omit the subscript $t$.
The unitary evolution operator can be obtained through time ordered integration over a Floquet period $T$: 
\begin{equation}
\hat{U}(T) = \mathcal{T}\left\{\text{exp}\left[-\frac{i}{\hbar}\int^{T}_{0} dt  \hat{H}(\rfph(t), \Omrf(t), \Omega_+, \Omega_-)\ \right] \right\}
\end{equation}
where the periodic time dependent Hamiltonian $\hat{H}(t)$ is explicitly controlled by the parameters of the dressing fields. 
We directly obtained $\hat{U}(T)$ using the Trotter decomposition, giving Floquet Hamiltonian $\hat H^{\rm F}$.

\section{Experiment}

Our experiment uses large $\Omega_\pm$ where its lowest two bands approximate the two-band tight binding SSH model.
A second important consequence of the deep lattice was a large energy gap to the excited bands, making their excitation less relevant for our Floquet drive.

In these deep 1D spin-dependent lattices, the maximally localized Wannier orbitals are localized very near the minima of the adiabatic potential, and as is explicit in Fig. 1 of the main text become highly magnetized.
We use this fact to extract the information of occupancy of the sites with high fidelity.

\subsection{Loading}

Our loading procedure begins with a BEC in the state $\ket{m_F = -1}$ . We detuned the bias field in $5\ \rm{ms}$ by $-100\ \rm{kHz}$ from the rf resonant frequency, holding for $5\ \rm{ms}$ for stabilization, and then exponentially ramp the Raman and RF fields on in $2.5\ {\rm ms}$ with $\rfph=0$ or $\pi$, while the bias field simultaneously ramps back to rf resonance. This adiabatic loading prepares atoms in the ground state of the lattice.

In this lattice configuration the sublattice sites are tilted by $\Delta \approx 6 \Er$ (much larger than $J$ or $J^\prime$), and the resulting ground state is highly polarized with $|m_x| > 0.99$ by simulation, which also agrees with experimental observations. 

This high polarization indicates the purity of initial state localized in either of the sites per cell. We then rapidly switch $\rfph$ to $\pm\pi/2$ (giving the degenerate sublattice sites described by the SSH model), populating either of the $\ket{\uparrow,\downarrow}$ states. 

We note that the localized states in the tilted lattice are slightly different than the targeted $\ket{\uparrow,\downarrow}$ states, for the local harmonic profiles of the lattice sites are largely stationary but nevertheless changeable in this deep lattice as $\rfph$ hops by $\pm \pi/2$. This could also be interpreted that the ground state of a tilted lattice has some mixture of higher bands in a degenerate lattice excluding the lowest two bands (which almost entirely constitutes the $\ket{\uparrow,\downarrow}$ states) when projected onto the latter basis. Our simulation indicates that this sudden jump has a small  $\approx 7~\%$ probability of excitation into excited bands, highly consistent with the experimental data. 

This sudden switching protocol was implemented owing to a limitation of our direct digital synthesis devices, and could be easily rectified in future experiments. 

\subsection{Calibration}

We calibrated the lattice parameters by conducting three separate Rabi pulsing experiments~\cite{lu2016geometrical}, first for resonant rf (giving $\Omrf$), and then for both Raman transitions (giving $\Omega_+$ and $\Omega_-$).
We then observed the spin evolution under the full Hamiltonian, and fit with $\rfph$ as the only free parameter.
This provided a course calibration of $\rfph$ accurate to a few degrees.
$\rfph$ was then calibrated with increased precision by following our loading procedure to produce a polarized state in a dimerized lattice and tuning $\rfph$ to maximize the amplitude and period of the resulting tunneling oscillations.
We confirmed that our full Floquet model accurately describes a Floquet experiment in which $\rfph$ changed by $\pi$ every half cycle, with Raman coupling applied.

We used an rf mixer to control the magnitude and sign of the rf field (allowing the required $\pi$ phase changes).
Real mixers and rf amplifiers are non-linear.
We performed preliminary calibrations of this nonlinearity off-line by commanding a range of driving levels and measuring the resulting rf amplitude.
We then fine tuned this calibration by performing rf-only Floquet experiments with the Tukey switching profiles (see below) commanded, and fit the resulting time evolution to a model including the mixer's non-linearity expressed as a polynomial expansion, with the polynomial coefficients as free parameters.
The fitted nonlinear coefficients were included in the full numerical simulation.

\subsection{Model parameters}

We choose $\rfph, \Omrf, \Omega_+, \Omega_-$ so that the dispersion of the two lowest bands is well described by the SSH model and fit to the SSH dispersion to obtain values for $J$ and $J^\prime$.
As shown in Fig.~2(a) unwanted tunneling can be made neglegible except during switchings, at which time $J$ and $J^\prime$ are simultaneously non-zero with a magnitude of about $1/5$ of the full tunneling strength.
Additionally, we computed the band gap between the two lowest bands to be $\Delta E_{0,1} \approx 2 J_0 \approx 0.6\ \Er$ and the energy spacing to the next higher bands to be $\Delta E_{0/1, n} \gtrapprox 6 \Er$.

\subsection{Timing}

The idealized protocol switches between configurations I and II instantly, however, as noted above each such switch would excited higher bands.  
Formally this creates crossings between the desired ground-band Floquet eigenstates and those associated with higher bands.
Because our Floquet protocols switch many times, such excitation/coupling is not acceptable.

Our experiment smoothed the switching behavior to prevent this behavior.
We empirically found the Tukey waveform with cosine edge fraction $\alpha = 0.3$ to be highly effective, giving $2\pi/T \approx \Delta E_{0,1} < 2\pi/(\alpha T) < \Delta E_{0/1, n}$.
As seen in the fully numerical spectra in Fig. 1, this choice had near linear bands with no gaps on the scale of $J_0$.

\end{document}